\documentclass[12pt,reqno]{article}
\usepackage{latexsym}
\usepackage{amsmath}
\usepackage[final]{pdfpages} 
\usepackage{graphicx}
\usepackage{amssymb}
\usepackage{euscript}
\usepackage{eufrak}
\usepackage{cancel}
\usepackage{fancyhdr}
\makeatletter\newlength{\earraycolsep}
\newlength{\textlarg}

\renewcommand{\thefootnote}{\alph{footnote}}
\begin{document}
\pagestyle{myheadings}

\title{Regge symmetry of  6-j or super 6-jS symbols: \\a re-analysis with partition properties\\}
\author{\tt Lionel Br\'{e}hamet \thanks {email: brehamet.l@orange.fr} \\ 
Research scientist \hspace{0.38em} 
\\ {France \hspace{4.92em} }
\date{}}
\maketitle

\begin{abstract}
\noindent 
It is shown that the five Regge transformations act as a spectrometric splitter on any 6-j symbol. Four unknown partitions are brought out: 
S4(0), S4(1), S(2) and S4(5).
They are stable subsets, with well defined parameters depending only on triangles and quadrangles. These findings 
are easily generalized to super 6-jS symbols, properly labelled by their own parity alpha, beta, gamma. 
Super Regge symmetry is reduced only for beta where S4(2), S4(5) vanish. In addition, all tools for computing exact values of any 6-jS are provided.
[0.2em]
\end{abstract}

\renewcommand{\thefootnote}{\arabic{footnote}}
\setcounter{footnote} {0}

\begin{flushleft}
\newpage
			\section{Introduction}\label{Intro}
Thanks to Regge \cite{Regge}, new symmetries became known since 1959 and relate to Racah-Wigner $n$-$j$ symbols. 
This is quite well referenced in standard books on Quantum Mechanics and Angular Momentum, like  \cite{Bied.Louck1}.
These symmetries generate new quantal triangles whose features have {\tt never} been fully investigated as will be done in this paper.

			\section{Recalls about $6$-$j$ or $6$-$j^{S}$ symbol}\label{GeneralParaneters}
\renewcommand{\theequation}{\ref{GeneralParaneters}.\arabic{equation}}
\setcounter{equation}{0}
Their numerical values and other properties are determined by seven non independent parameters 
$p_1,p_2,p_3,p_4$ (triangles) and $q_1,q_2,q_3$ (quadrangles) defined below:\\

\setlength{\unitlength}{0.2in}
\begin{picture}(30,3)
\put(7,2){\makebox(0,0){\scriptsize \bf  The four quantal triangles of any $6$-$j$}}
\put(24,2){\makebox(0,0){\scriptsize \bf The three different pairs of columns of any $6$-$j$}}
\put(0,0){\circle*{.125}} \put(1,1){\circle*{.125}} \put(2,1){\circle*{.125}}
\put(0,1){\circle{.25}} \put(1,0){\circle{.25}} \put(2,0){\circle{.25}}
\put(0,1){\line(1,-1){1}} \put(1,0){\line(1,0){1}}
\put(4,1){\circle*{.125}} \put(5,0){\circle*{.125}} \put(6,1){\circle*{.125}}
\put(4,0){\circle{.25}} \put(5,1){\circle{.25}} \put(6,0){\circle{.25}}
\put(4,0){\line(1,1){1}} \put(5,1){\line(1,-1){1}}
\put(8,1){\circle*{.125}} \put(9,1){\circle*{.125}} \put(10,0){\circle*{.125}}
\put(8,0){\circle{.25}} \put(9,0){\circle{.25}} \put(10,1){\circle{.25}}
\put(8,0){\line(1,0){1}} \put(9,0){\line(1,1){1}}
\put(12,0){\circle*{.125}} \put(13,0){\circle*{.125}} \put(14,0){\circle*{.125}}
\put(12,1){\circle{.25}} \put(13,1){\circle{.25}} \put(14,1){\circle{.25}}
\put(12,1){\line(1,0){1}} \put(13,1){\line(1,0){1}}

\put(18,0){\circle*{.125}} \put(19,0){\circle{.25}} \put(20,0){\circle{.25}}
\put(18,1){\circle*{.125}} \put(19,1){\circle{.25}} \put(20,1){\circle{.25}}
\put(19,1){\line(0,-1){1}} \put(20,1){\line(0,-1){1}}

\put(23,0){\circle{.25}} \put(24,0){\circle*{.125}} \put(25,0){\circle{.25}}
\put(23,1){\circle{.25}} \put(24,1){\circle*{.125}} \put(25,1){\circle{.25}}
\put(23,1){\line(0,-1){1}} \put(25,1){\line(0,-1){1}}

\put(28,0){\circle{.25}} \put(29,0){\circle{.25}} \put(30,0){\circle*{.125}}
\put(28,1){\circle{.25}} \put(29,1){\circle{.25}} \put(30,1){\circle*{.125}}
\put(28,1){\line(0,-1){1}} \put(29,1){\line(0,-1){1}}

\put(1,-1){\makebox(0,0){$p_1$}} \put(5,-1){\makebox(0,0){$p_2$}} \put(9,-1){\makebox(0,0){$p_3$}} \put(13,-1){\makebox(0,0){$p_4$}}
\put(19,-1){\makebox(0,0){$q_1$}} \put(24,-1){\makebox(0,0){$q_2$}} \put(29,-1){\makebox(0,0){$q_3$}}
\end{picture}\par \vspace{2em}
The notations adopted are:  $p_i ,i=[1,4]$ and  $q_k ,k=[1,3]$, where $p_i$ is the sum of the values of the three circled spins just above
$p_i$ in the diagrams. In the same way, $q_k$ is the sum of the values of  the four circled spins above $q_k$. 

						\subsection{Analytic formula for standard $6$-$j$ symbols}\label{StandardFormula}
\renewcommand{\theequation}{\ref{StandardFormula}\alph{equation}}
\setcounter{equation}{0}
 Expression of a  $6$-$j$ symbol is well known
 \cite[p. 99] {Edmonds}, however we leave out the traditional notation with triangles  $\bigtriangleup$ for adopting here a {\tt re-transcription in terms of $p_i,q_k$}:\\
\begin{equation} \label{eq:General6j1}
\left\{ \begin{array}{ccc} J_1 &J_2 & J_{3} \\ j_1 & j_2 & j_3 \end{array} \right\}= 
\left[ \frac{\prod_{k=1}^{k=3}\prod_{i=1}^{i=4}(q_k-p_i)!}{\prod_{i=1}^{i=4}(p_i+1)!}\right]^{\frac{1}{2}} 
 \begin{array}{c}{\displaystyle \sum_{z}} \frac{(-1)^{z}(z+1)! } 
{\prod_{i=1}^{i=4}(z-p_i)! \prod_{k=1}^{k=3}( q_k-z)!  }\, .\end{array}
\end{equation}


						\subsection{Analytic formula for super $6$-$j^{S}$ symbols}\label{SuperFormula}
\renewcommand{\theequation}{\ref{SuperFormula}\alph{equation}}
\setcounter{equation}{0}
Extension from $so(3)$ to super algebra $osp(1|2)$ was finalized in a paper dated $1993$ by Daumens {\it et al.} \cite{Daumensetal1} .
Then a standard $6$-$j$ symbol becomes a super $6$-$j^{S}$ symbol denoted by $\left\{ \begin{array}{ccc} J_1 &J_2 & J_{3} \\
j_1 & j_2 & j_3 \end{array} \right\}^{S}$ . Essential Racah-Wigner calculus properties were studied.\\
\hspace*{1em}{\tt However} no analytic formula was brought out for expressing a super \{$6$-$j$\}$^{S}$ symbol as a single summation over an integer $z$ like for $so(3)$.\\
In 2006 we succeeded in derive such a formula \cite{L.B.Nuov1.2006}. It was found that any \{$6$-$j$\}$^{S}$ should be labelled by an additional parity parameter
$\pi$=$\alpha,\beta,\gamma$ according to new properties of these symbols namely:  triangles $p_i$ can be integer or half-integer. A right notation will be \{$6$-$j$\}$_\pi^{S}$ .\\
Let us recall our definitions :
\begin{equation}   \label{eq:parities}
\left\{\begin{array}{l}
\pi=\alpha   \mbox{\hspace{2em}if\hspace{1em}}   \forall i\in [1,4] \mbox{\hspace{1em}} p_i  \mbox{\hspace{0.5em}integer}\\
\pi=\beta    \mbox{\hspace{2em}if\hspace{1em}} \forall i,j,k,l \hspace{0.2em}\in[1,4] \hspace{0.2em} (i\neq j \neq k\neq l)
\mbox{\hspace{1em}} p_i,p_j  \mbox{\hspace{0.5em}half-integer,} \mbox{\hspace{1em}} p_k,p_l  \mbox{\hspace{0.5em}integer}\\
\pi=\gamma   \mbox{\hspace{2em}if\hspace{1em}}   \forall i\in [1,4] \mbox{\hspace{1em}} p_i \mbox{\hspace{0.5em}half-integer} 
\end{array}\right. .
\end{equation}
In the case of parity $\boldsymbol{\beta}$, both integer triangles shall be denoted by $p,p'$,
both other half-integer by  $\overline{p}, \overline{p}'$. The single integer quadrangle is denoted by $q$, 
both other half-integer  by $\overline{q}, \overline{q}'$.\\ [0.2em]
\hspace*{1em}Here again we leave out our own notations with super triangles $\triangle^{S}$ for showing compact expressions.
Delimiter $[\;]$ around expressions  like $q_k-p_i, p_i+\frac{1}{2},q_k+\frac{1}{2}$ means 'integer part of'.
General formula of any \{$6$-$j$\}$_\pi^{S}$ thus may be written as follows:
\begin{equation} \label{eq:General6jS1}
\left\{ \begin{array}{ccc} J_1 &J_2 & J_{3} \\ j_1 & j_2 & j_3 \end{array} \right\}_{\pi}^{S}= 
(-1)^{  \scriptsize 4 \begin{array}{c}\displaystyle \sum_{k=1}^{k=3}\! J_k j_k \end{array}}\!\!
\left[ \frac{\prod_{k=1}^{k=3}\prod_{i=1}^{i=4}[q_k-p_i]!}{\prod_{i=1}^{i=4}[p_i+\frac{1}{2}]!}\right]^{\frac{1}{2}} 
\begin{array}{c} {\displaystyle  \! \sum_{z}} \frac{(-1)^{z}z! \Pi_{\pi}(z) } 
{\prod_{i=1}^{i=4}\left( z-[p_i+\frac{1}{2}]\right)! \prod_{k=1}^{k=3}
\left( [q_k+\frac{1}{2}]-z)\right)! }\, .\end{array}
\end{equation}

Monomials $\Pi_\pi(z)$ are of degree $0$ or $1$ in $z$: 
\begin{eqnarray}
\lefteqn{
\mbox{\hspace{13.7em}} \Pi_{\alpha}(z)= 1,   \label{eq:PiMonomS1}}\\
&\Pi_{\beta}(z)= \begin{array}{c}-z(\overline{q}+\overline{q}' -p-p'+1)+(\overline{q}+\frac{1}{2})(\overline{q}' +\frac{1}{2})- pp' \end{array},   \label{eq:PiPiMonomS2}\\
&\Pi_{\gamma}(z)=-z+2(J_1j_1+J_2j_2+J_3j_3)+(J_1+j_1+J_2+j_2+J_3+j_3)+\begin{array}{c}\frac{1}{2}\end{array}. \label{eq:PiPiMonomS3}
\end{eqnarray}
As it is obvious for $\pi=\alpha$ or $\beta$, all constant terms, factor of $z$ and $\Pi_{\alpha,\beta}(0) $, are integers.\\
For parity $\gamma$ however, a detailed check is necessary for realizing that $\Pi_{\gamma}(0) $ is integer also.\\
This is useful for computing $6$-$j^{S}$ numerical values, as we did it with program ({\tt superspins}).\\
{\tt Let us stress that our new expressions (\ref{eq:General6j1}), (\ref{eq:General6jS1})  in terms of $p$ and $q$ are far to be  an artifact}.
Indeed they potentially contain all new properties that will be highlighted in this article.

				\section{Regge symmetry of a $6$-$j$ symbol: a new formulation }\label{ReggeSect3}
\renewcommand{\theequation}{\ref{ReggeSect3}.\arabic{equation}}
\setcounter{equation}{0}
\hspace{1.5em}It was shown \cite{Regge} that linear transformations on the spins involved in any classical $6$-$j$ symbol
lead to additional symmetries, out of the $S_4$ tetrahedron symmetry.\\
On the left: the well known expressions, on the right or below: our formulation in terms of  $q$.
\begin{eqnarray}
\scriptsize
\left\{ \begin{array}{ccc} J_1 &J_2 & J_{3} \\
j_1 & j_2 & j_3 \end{array} \right\} &=
\scriptsize
\left\{ \begin{array}{ccc} J_1 & \frac{1}{2}(J_2+J_3+j_2-j_3)&  \frac{1}{2}(J_2+J_3+j_3-j_2)\\ [0.1em]
j_1 & \frac{1}{2}(J_2+j_3+j_2-J_3) &\frac{1}{2}(J_3+j_3+j_2-J_2) \end{array} \right\}&=
\scriptsize
\left\{ \begin{array}{ccc} J_1 &\frac{1}{2}q_1-J_2 & \frac{1}{2}q_1-J_3  \\ [0.1em]
j_1 &\frac{1}{2}q_1-j_2 &\frac{1}{2}q_1-j_3 \end{array} \right\}, \label{Reggetrsf1}\\ 
							&=
\scriptsize
\left\{ \begin{array}{ccc} \frac{1}{2}(J_1+J_3+j_1-j_3) & J_2 &  \frac{1}{2}(J_1+J_3+j_3-j_1)\\ [0.1em]
 \frac{1}{2}(j_1+J_1+j_3-J_3) &j_2 &\frac{1}{2}(j_3+j_1+J_3-J_1) \end{array} \right\}&=
\scriptsize
\left\{ \begin{array}{ccc} J_2 &\frac{1}{2}q_2-J_3 & \frac{1}{2}q_2-J_1  \\ [0.1em]
j_2 &\frac{1}{2}q_2-j_3 &\frac{1}{2}q_2-j_1 \end{array} \right\}, \label{Reggetrsf2}\\
							&=
\scriptsize
\left\{ \begin{array}{ccc} \frac{1}{2}(J_1+J_2+j_1-j_2) &  \frac{1}{2}(J_2+J_1+j_2-j_1) & J_3\\ [0.1em]
 \frac{1}{2}(j_1+J_1+j_2-J_2) & \frac{1}{2}(j_2+j_1+J_2-J_1)&j_3  \end{array} \right\}&=
\scriptsize
\left\{ \begin{array}{ccc} J_3 &\frac{1}{2}q_3-J_1 & \frac{1}{2}q_3-J_2  \\ [0.1em]
j_3 &\frac{1}{2}q_3-j_1 &\frac{1}{2}q_3-j_2 \end{array} \right\}\label{Reggetrsf3}.
\end{eqnarray}
\begin{eqnarray}
\scriptsize
\left\{ \begin{array}{ccc} J_1 &J_2 & J_{3} \\
j_1 & j_2 & j_3 \end{array} \right\}&=
\scriptsize
\left\{ \begin{array}{ccc} \frac{1}{2}(J_2+j_2+J_3-j_3) &  \frac{1}{2}(J_1+j_3+J_3-j_1) &  \frac{1}{2}(J_1+j_1+J_2-j_2)\\ [0.1em]
 \frac{1}{2}(J_2+j_3+j_2-J_3) & \frac{1}{2}(J_3+j_1+j_3-J_1)&\frac{1}{2}(J_1+j_1+j_2-J_2) \end{array} \right\} \label{Reggetrsf4}\\ 
&=  
\scriptsize
\left\{ \begin{array}{ccc}\frac{1}{2}q_1-j_3  &\frac{1}{2}q_2-j_1 & \frac{1}{2}q_3-j_2  \\ [0.1em]
\frac{1}{2}q_1-J_3  &\frac{1}{2}q_2-J_1 & \frac{1}{2}q_3-J_2   \end{array} \right\}\nonumber, \\ [1em]
&=
\scriptsize
\left\{ \begin{array}{ccc} \frac{1}{2}(J_2+j_3+J_3-j_2) &  \frac{1}{2}(J_1+j_1+J_3-j_3) &  \frac{1}{2}(J_1+j_2+J_2-j_1)\\ [0.1em]
\frac{1}{2}(J_3+j_3+j_2-J_2) & \frac{1}{2}(J_1+j_1+j_3-J_3)&\frac{1}{2}(J_2+j_1+j_2-J_1) \end{array} \right\} \label{Reggetrsf5}\\ 
&= 
\scriptsize
\left\{ \begin{array}{ccc}\frac{1}{2}q_1-j_2  &\frac{1}{2}q_2-j_3 & \frac{1}{2}q_3-j_1  \\ [0.1em]
\frac{1}{2}q_1-J_2  &\frac{1}{2}q_2-J_3 & \frac{1}{2}q_3-J_1   \end{array} \right\}\nonumber . 
\end{eqnarray}

By using a cyclic index notation, the transformations can be written under a compact form:
\begin{equation}\label{eq:ReggeA0}\scriptsize
\left\{ \begin{array}{ccc} J_1 &J_2 & J_3 \\ j_1 & j_2 & j_3 \end{array} \right\}=
\left\{ \begin{array}{ccc} J_l &\frac{1}{2}q_l-J_m & \frac{1}{2}q_l-J_n  \\ [0.1em]
j_l &\frac{1}{2}q_l-j_m &\frac{1}{2}q_l-j_n \end{array} \right\} \quad\; \normalsize
 l\in[1,3] \;   l,m,n \;  \mbox{\hspace{0.1em}cyclic\;on  } \hspace{0.1em}[1,2,3] .
\end{equation} 

With the same cyclic notations, one may write the remaining Regge transformations as
\begin{eqnarray}\label{eq:ReggeA1}
\scriptsize
\left\{ \begin{array}{ccc} J_1 &J_2 & J_3 \\ j_1 & j_2 & j_3 \end{array} \right\}& =
\scriptsize
\left\{ \begin{array}{ccc}\frac{1}{2}q_l-j_m  &\frac{1}{2}q_m-j_n & \frac{1}{2}q_n-j_l  \\ [0.1em]
\frac{1}{2}q_l-J_m  &\frac{1}{2}q_m-J_n & \frac{1}{2}q_n-J_l   \end{array} \right\}& =
\scriptsize
 \left\{ \begin{array}{ccc}\frac{1}{2}q_l-j_n  &\frac{1}{2}q_m-j_l & \frac{1}{2}q_n-j_m  \\ [0.1em]
\frac{1}{2}q_l-J_n  &\frac{1}{2}q_m-J_l & \frac{1}{2}q_n-J_m   \end{array} \right\}.
\end{eqnarray}

Note some useful relations between triangles and quadrangles:
\begin{equation}
q_l=p_m+p_n-2j_l \, , \; q_l=p_l+p_4-2J_l  \;\,  \mbox{and}\;
\begin{array}{c} \sum_{k=1}^{k=3}\, q_k \end{array} \!\!= \!\!\begin{array}{c} \sum_{l=1}^{l=4}\, p_l \end{array}.
\end{equation}
As $2j_l$ is integer any standard $6$-$j$ symbol has all triangles $p$ integer, thus any $q_l$ is integer. \\ 
\hspace*{1em} Actually alternative formulations of Regge transformations are available. As triangles $p$ and quadrangles $q$ are essential keys 
of $6$-$j$ symbols algebraic formula, we are going to use them for looking exactly how their spins are changed by these transformations. 
Exactly 12 (non trivial) different algebraic expressions are needed and contained only in the first three equations (\ref{Reggetrsf1}-\ref{Reggetrsf3}). 
For each of them, in terms of $p$ or $q$, one has 3 equivalent formulations.
It results in four basic equations written with a cyclic notation on $l,m,n$ cyclic on $[1,2,3]$.
\begin{eqnarray}
\begin{array}{cccc} \label{eq:Alteq1}
\frac{1}{2}(J_l+J_m+j_l-j_m)=&J_l+\frac{1}{2}(p_m-p_l) =&j_l+\frac{1}{2}(p_4-p_n)= &J_m+\frac{1}{2}(q_m-q_l), \end{array}\\
\begin{array}{cccc}  \label{eq:Alteq2}
\frac{1}{2}(j_l+j_m+J_l-J_m)=&j_l+\frac{1}{2}(p_l-p_m)=&J_l-\frac{1}{2}(p_4-p_n) = &j_m+\frac{1}{2}(q_m-q_l), \end{array}\\  [0.4em]
\begin{array}{cccc}  \label{eq:Alteq3}
\frac{1}{2}(J_m+J_l+j_m-j_l) =&J_m+\frac{1}{2}(p_l-p_m)=&j_m+\frac{1}{2}(p_4-p_n) = &J_l+\frac{1}{2}(q_l-q_m), \end{array}\\ 
\begin{array}{cccc}  \label{eq:Alteq4}
\frac{1}{2}(j_m+j_l+J_m-J_l)=&j_m+\frac{1}{2}(p_m-p_l)=&J_m-\frac{1}{2}(p_4-p_n) = &j_l+\frac{1}{2}(q_l-q_m). \end{array}
\end{eqnarray}


						\subsection{Matrix representation of Regge transformations}\label{ReggeMatrix}
\renewcommand{\theequation}{\ref{ReggeMatrix}\alph{equation}}
\setcounter{equation}{0}
Consider a 6-dimensional vector space where any $6$-$j$ is  a vector $\mathbf{J}$ with 6-components, $J_1,J_2,J_3,j_1,j_2,j_3$, on a canonical basis. 
Then the five Regge transformations (\ref{Reggetrsf1}-\ref{Reggetrsf5}) are represented  by five  $6\times6$ matrix $\mathbf{R}_\kappa$, $\kappa \in [1,5]$.
according to the following scheme: 

\begin{equation}
\scriptsize
\mathbf{J'}=\mathbf{R}_{\kappa} \mathbf{J}. \normalsize
\end{equation}
\begin{eqnarray}\scriptsize
\begin{array}{c}
\mathbf{J} \\ [0.2em]
\left[ \begin{array}{c}J_1\\J_2\\J_3\\j_1\\j_2\\j_3
\end{array}\right],
\end{array}
\begin{array}{c}
\mathbf{J'} \\ [0.2em]
\left[ \begin{array}{c}J'_1\\J'_2\\J'_3\\j'_1\\j'_2\\j'_3
\end{array}\right],
\end{array}
\begin{array}{c}
\mathbf{R}_1 \\ [0.2em]
\left( \begin{array}{rrrrrr}
1 & 0 & 0 \;\,& 0 & 0 & 0 \\ [0.1em]
0 & \frac{1}{2} & \frac{1}{2} \;\,& 0 & \frac{1}{2} & -\frac{1}{2} \\ [0.1em]
0 & \frac{1}{2} & \frac{1}{2} \;\,& 0 & -\frac{1}{2} & \frac{1}{2} \\  [0.1em]
0 & 0 & 0 \;\,& 1 & 0 & 0 \\  [0.1em]
0 & \frac{1}{2} & -\frac{1}{2}\;\,& 0 & \frac{1}{2} & \frac{1}{2} \\  [0.1em]
0 & -\frac{1}{2} & \frac{1}{2} \;\,& 0 & \frac{1}{2} & \frac{1}{2}
\end{array}\right) 
\end{array}, 
\begin{array}{c}
\mathbf{R}_2 \\ [0.2em]
\left( \begin{array}{rrrrrr}
 \frac{1}{2} & \;\,0  &  \frac{1}{2}&  \frac{1}{2} \;\,& 0 & - \frac{1}{2} \\  [0.1em]
0 & 1 & 0 & 0 \;\,& 0 & 0 \\  [0.1em]
 \frac{1}{2} & 0 &  \frac{1}{2} & - \frac{1}{2} \;\,& 0 &  \frac{1}{2} \\  [0.1em]
 \frac{1}{2} & 0 & - \frac{1}{2} &  \frac{1}{2} \;\,& 0 &  \frac{1}{2} \\  [0.1em]
0 & 0 & 0 & 0 \;\,& 1 & 0 \\  [0.1em]
- \frac{1}{2} & 0 &  \frac{1}{2} &  \frac{1}{2} \;\,& 0 &  \frac{1}{2}
\end{array}\right) 
\end{array}, 
\end{eqnarray}
\begin{eqnarray}\scriptsize
\begin{array}{c}
\mathbf{R}_3 \\ [0.2em]
\left( \begin{array}{rrrrrr}
 \frac{1}{2} &  \frac{1}{2}  \;\,& 0&  \frac{1}{2} & - \frac{1}{2} \;\,& 0 \\ [0.1em]
 \frac{1}{2} &  \frac{1}{2} \;\,& 0 & - \frac{1}{2} &  \frac{1}{2} \;\,& 0 \\ [0.1em]
0 & 0  \;\,& 1 &0 & 0 \;\,& 0 \\ [0.1em]
 \frac{1}{2} & - \frac{1}{2} \;\,& 0 &  \frac{1}{2} &  \frac{1}{2} \;\,& 0 \\ [0.1em]
- \frac{1}{2} &  \frac{1}{2} \;\,& 0 &  \frac{1}{2} &  \frac{1}{2} \;\,& 0 \\ 
0 & 0  \;\,& 0  &  0& 0 \;\,& 1
\end{array}\right)
\end{array},
\begin{array}{c}
\mathbf{R}_4 \\ [0.2em]
\left( \begin{array}{rrrrrr}
0 &  \frac{1}{2} &  \frac{1}{2} & 0 &  \frac{1}{2} & - \frac{1}{2} \\ [0.1em]
 \frac{1}{2} & 0 &  \frac{1}{2} & - \frac{1}{2} & 0 &  \frac{1}{2} \\ [0.1em]
 \frac{1}{2} &  \frac{1}{2} & 0 &  \frac{1}{2} & - \frac{1}{2} & 0 \\ [0.1em]
0 &  \frac{1}{2}& - \frac{1}{2}& 0 &  \frac{1}{2} &  \frac{1}{2} \\ [0.1em]
- \frac{1}{2} & 0 &  \frac{1}{2} &  \frac{1}{2} & 0 &  \frac{1}{2} \\ [0.1em]
 \frac{1}{2} & - \frac{1}{2} & 0 &  \frac{1}{2} &  \frac{1}{2} & 0
\end{array}\right) 
\end{array}, 
\begin{array}{c}
\mathbf{R}_5 \\ [0.2em]
\left( \begin{array}{rrrrrr}
0 & \frac{1}{2}  & \frac{1}{2} & 0 & -\frac{1}{2} & \frac{1}{2} \\ [0.1em]
\frac{1}{2} & 0 & \frac{1}{2} & \frac{1}{2} & 0 & -\frac{1}{2} \\ [0.1em]
\frac{1}{2} & \frac{1}{2} & 0 & -\frac{1}{2} & \frac{1}{2} & 0 \\ [0.1em]
0 & -\frac{1}{2} & \frac{1}{2} & 0 & \frac{1}{2} & \frac{1}{2} \\ [0.1em]
\frac{1}{2} & 0 & -\frac{1}{2} & \frac{1}{2}& 0 & \frac{1}{2} \\ [0.1em]
-\frac{1}{2} & \frac{1}{2} & 0 & \frac{1}{2} & \frac{1}{2} & 0
\end{array}\right) 
\end{array} .
\end{eqnarray}
\normalsize
All these matrices are invertible, diagonalizable and have various properties like 
\begin{equation}
(\mathbf{R}_1)^2 =(\mathbf{R}_2)^2 =(\mathbf{R}_3)^2  = \mathit{I_d}, \quad 
det(\mathbf{R}_1)= det(\mathbf{R}_2)= det(\mathbf{R}_3)=-1. 
\end{equation}
\begin{equation}
\mathbf{R}_1 \times \mathbf{R}_2 =  \mathbf{R}_2 \times \mathbf{R}_3 =  \mathbf{R}_3 \times \mathbf{R}_1
\quad \mbox{and} \quad \mathbf{R}_2 \times \mathbf{R}_1 =  \mathbf{R}_3 \times \mathbf{R}_2 =  \mathbf{R}_1 \times \mathbf{R}_3.
\end{equation}
\begin{equation}
\mathbf{R}_4\times \mathbf{R}_5 = \mathbf{R}_5\times \mathbf{R}_4=  \mathit{I_d}, \quad det(\mathbf{R}_4)= det(\mathbf{R}_5)=1.
\end{equation}
 $(\begin{array}{cccccc}-1,&1,&1,&1,&1,&1 \end{array}$) and 
$\left( \begin{array}{cccccc}\frac{(i \sqrt{3}-1)}{2},& \frac{(i \sqrt{3}-1)}{2},& 
-\frac{(i \sqrt{3}+1)}{2},& -\frac{(i \sqrt{3}+1)}{2},& 1,& 1 \end{array}\right)$ are respectively the eigenvalues of
 $\mathbf{R}_1,\mathbf{R}_2,\mathbf{R}_3$  and $\mathbf{R}_4, \mathbf{R}_5$. As said above $\mathbf{R}_4$ or $\mathbf{R}_5$
contains two rows of $\mathbf{R}_1$, two rows of $\mathbf{R}_2$ and two rows of $\mathbf{R}_3$. 
It can be checked also that all Regge transformations leave invariant the forms $J_1j_1$+$J_2j_2$+$J_3j_3$ and $J_1$+$j_1$+$J_2$+$j_2$+$J_3$+$j_3$
= $\begin{array}{c}\frac{1}{2} \sum_{k=1}^{k=3}\, q_k \end{array}$.\\
If symbolically  $\|  \mathbf{J}\|$ stands for the numerical value of {\scriptsize $ \mathbf{J}= \left\{ \begin{array}{ccc} J_1 &J_2 & J_{3} \\
j_1 & j_2 & j_3 \end{array} \right\}$}, one may say that  $\|  \mathbf{J}\|$ is left invariant by any $\mathbf{R}_\kappa$.



						\subsection{Features of $6$-$j$ symbols generated by Regge transformations}\label{ReggeSect3prime}
\renewcommand{\theequation}{\ref{ReggeSect3prime}\alph{equation}}
\setcounter{equation}{0}
To our knowledge, no real analysis was performed {\tt with a view to algebraic partition}.\\
Analogously with the $S_4$  tetrahedron permutations, let us introduce some definitions and notations easily understandable.
Any $6$-$j$ symbol has 24 possible distinct aspects (itself included). This set is denoted by $\tt{S}_4$. Each of the five equations
(\ref{Reggetrsf1}-\ref{Reggetrsf5}) is a Regge transformation that we have numbered  from $1$ to $5$ . That can be diagrammed as follows:
\begin{equation}
\{6\mbox{-}j\}\in {\tt{S}}_4 \,\stackrel{Regge_\kappa}\longrightarrow \{6\mbox{-}j\}^{{\cal{R}}_\kappa}\, \in {\tt{S}}_4 ^{{\cal{R}}_\kappa},
\quad \kappa \in [1,5] .
\end{equation}
$\forall  \lambda \neq \kappa$ either sets are the same  i.e.  
$ {\tt{S}}_4 ^{{\cal{R}}_\kappa}
\cap {\tt{S}}_4 ^{{\cal{R}}_\lambda}={\tt{S}}_4 ^{{\cal{R}}_\kappa} $, 
or disjoint i.e. $ {\tt{S}}_4 ^{{\cal{R}}_\kappa}
\cap {\tt{S}}_4 ^{{\cal{R}}_\lambda}=\emptyset $. 
\\ \vspace{0.5em}

{ \sl{A priori}},  $6$ \tt{disjoint} \rm  sets should be defined, namely ${\tt{S}}_4 (0)$, ${\tt{S}}_4 (1)$, ${\tt{S}}_4 (2)$, ${\tt{S}}_4 (3)$,
${\tt{S}}_4 (4)$, ${\tt{S}}_4 (5)$.
\begin{eqnarray}\label{eq:DisjtEqs}
\begin{array}{l}
{\tt{S}}_4 (0)= \left\{\{6\mbox{-}j\}\;| \, {\tt{S}}_4 ^{{\cal{R}}_5}
= {\tt{S}}_4 ^{{\cal{R}}_4}= {\tt{S}}_4 ^{{\cal{R}}_3}
= {\tt{S}}_4 ^{{\cal{R}}_2}= {\tt{S}}_4 ^{{\cal{R}}_1}= {\tt{S}}_4 \right\},\\[0.1em]
{\tt{S}}_4 (1)= \left\{\{6\mbox{-}j\}\;| \, \exists \; \mbox{only one $\kappa$ }|  {\tt{S}}_4 ^{{\cal{R}}_\kappa}
\cap {\tt{S}}_4 =\emptyset
\right\},\\[0.2em]
{\tt{S}}_4 (2)= \left\{\{6\mbox{-}j\}\;| \, \exists \; \mbox{only two $\kappa,\lambda$ }| \; 
{\tt{S}}_4 ^{{\cal{R}}_\kappa}\cap {\tt{S}}_4 ^{{\cal{R}}_\lambda}\cap {\tt{S}}_4 =\emptyset
\right\},\\[0.2em]
{\tt{S}}_4 (3)= \left\{\{6\mbox{-}j\}\;| \, \exists \; \mbox{only three $\kappa,\lambda,\mu$ }| \; 
{\tt{S}}_4 ^{{\cal{R}}_\kappa} \cap {\tt{S}}_4 ^{{\cal{R}}_\lambda} \cap{\tt{S}}_4 ^{{\cal{R}}_\mu} \cap {\tt{S}}_4 =\emptyset
\right\},\\[0.2em]
{\tt{S}}_4 (4)= \left\{\{6\mbox{-}j\}\;| \, \exists \; \mbox{only four $\kappa,\lambda,\mu,\nu$  }| \; 
{\tt{S}}_4 ^{{\cal{R}}_\kappa}\cap {\tt{S}}_4 ^{{\cal{R}}_\lambda}\cap {\tt{S}}_4 ^{{\cal{R}}_\mu}
\cap{\tt{S}}_4 ^{{\cal{R}}_\nu } \cap {\tt{S}}_4 =\emptyset \right\},\\
{\tt{S}}_4 (5)= \left\{\{6\mbox{-}j\}\,\;| \,  {\tt{S}}_4 ^{{\cal{R}}_5}
\cap {\tt{S}}_4 ^{{\cal{R}}_4}\cap {\tt{S}}_4 ^{{\cal{R}}_3}
\cap {\tt{S}}_4 ^{{\cal{R}}_2}\cap {\tt{S}}_4 ^{{\cal{R}}_1}\cap{\tt{S}}_4  =\emptyset\right\}.
\end{array}
\end{eqnarray}
${\cal{R}}_{all}$ will denote the set of the five Regge transformations acting on a given \{$6$-$j$\} symbol.\\
\hspace*{1em}Consider an example related to ${\tt{S}}_4 (2)$. ${\cal{R}}_{all}$ applied to
$\scriptsize \left\{ \begin{array}{ccc} $9$ & $8$ & $6$ \\  $3/2$ & $9/2$ & $13/2$ \end{array} \right\}_{0} $
 generates five  \{$6\mbox{-}j$\}:	
\begin{equation}\label{eq:List5}\scriptsize
\left\{ \begin{array}{ccc} $9$ & $9/2$ & $13/2$ \\  $3/2$ & $8$ & $6$ \end{array} \right\}^{{\cal{R}}_1}, 
\left\{ \begin{array}{ccc} $8$ & $11/2$ & $5/2$ \\  $9/2$ & $5$ & $10$ \end{array} \right\}^{{\cal{R}}_2}, 
\left\{ \begin{array}{ccc} $6$ & $5/2$ & $7/2$ \\  $13/2$ & $10$ & $7$ \end{array} \right\}^{{\cal{R}}_3}, 
\left\{ \begin{array}{ccc} $6$ & $10$ & $7$ \\  $13/2$ & $5/2$ & $7/2$ \end{array} \right\}^{{\cal{R}}_4}, 
\left\{ \begin{array}{ccc} $8$ & $5$ & $10$ \\  $9/2$ & $11/2$ & $5/2$ \end{array} \right\}^{{\cal{R}}_5}.
\normalsize 
\end{equation}
As $\exists \, s \in S_4\, | \scriptsize \left\{ \begin{array}{ccc} $9$ & $9/2$ & $13/2$ \\  $3/2$ & $8$ & $6$ \end{array} \right\}^{{\cal{R}}_1}
\stackrel{s}\rightarrow \scriptsize \left\{ \begin{array}{ccc} $9$ & $8$ & $6$ \\  $3/2$ & $9/2$ & $13/2$ \end{array} \right\}_{0}$,
we remove $\cancel{\scriptsize \left\{ \begin{array}{ccc} $9$ & $9/2$ & $13/2$ \\  $3/2$ & $8$ & $6$ \end{array} \right\}^{{\cal{R}}_1}}$
from the list (\ref{eq:List5}),
the same holds for $\cancel{\scriptsize \left\{ \begin{array}{ccc} $8$ & $5$ & $10$ \\  $9/2$ & $11/2$ & $5/2$ \end{array} \right\}^{{\cal{R}}_5}}\normalsize$
with $\scriptsize \left\{ \begin{array}{ccc} $8$ & $11/2$ & $5/2$ \\  $9/2$ & $5$ & $10$ \end{array} \right\}^{{\cal{R}}_2}\normalsize$, and
$\cancel{\scriptsize \left\{ \begin{array}{ccc} $6$ & $10$ & $7$ \\  $13/2$ & $5/2$ & $7/2$ \end{array} \right\}^{{\cal{R}}_4}}\normalsize$
with $\scriptsize \left\{ \begin{array}{ccc} $6$ & $5/2$ & $7/2$ \\  $13/2$ & $10$ & $7$ \end{array} \right\}^{{\cal{R}}_3}\normalsize$. 
Thus, after last convenient\footnote {means here with spins arranged according to the entries listed in Table like that of Rotenberg and al. \cite{Rot}.}
$S_4$ rearrangements, it remains only three representative  \{$6$-$j$\} of the set  ${\tt{S}}_4 (2)$ in our example, numbered from $0$ to $2$:
\begin{equation}\scriptsize
\left\{ \begin{array}{ccc} $9$ & $8$ & $6$ \\  $3/2$ & $9/2$ & $13/2$ \end{array} \right\}_{0} 
\; \mbox{\scriptsize{\&}}\left\{ \begin{array}{ccc} $10$ & $8$ & $5$ \\ $5/2$ & $9/2$ & $11/2$ \end{array} \right\}^{({\cal{R}}_2)}_{1}
\; \mbox{\scriptsize{\&}}\left\{ \begin{array}{ccc} $10$ & $7$ & $6$ \\ $5/2$ & $7/2$ & $13/2$ \end{array} \right\}^{({\cal{R}}_3)}_{2}
{\scriptstyle = -\frac{1}{2}\sqrt{\frac{23}{5\!\times\! 7\!\times \!13\! \times \!17}}}\; .\normalsize
\end{equation}
The operation realized for reducing the list  (\ref{eq:List5}) acts as a filter that we denote by $(S_{4}\,\mbox{\scriptsize filter})$.\\
{\tt Definition} of ${\cal{R}}_{egge}^{*}$:\\
\begin{equation}
{\cal{R}}_{egge}^{*}= (S_{4}\,\mbox{\scriptsize filter})\circ {\cal{R}}_{all}.
\end{equation}
More generally, $\forall$ \{$6\mbox{-}j$\}$_{0}\in {\tt{S}}_4 (2)$, ${\cal{R}}_{egge}^{*}$ generates two
'different` \{$6$-$j$\} symbols, namely \{$6\mbox{-}j$\}$_{1}$ and \{$6\mbox{-}j$\}$_{2}$. Numerical values are equal,
but their triangles are not the same. One has:
\begin{equation}
\begin{array}{c}
\{6\mbox{-}j\}_{0}\stackrel{{\cal{R}}_{egge}^{*}}\longrightarrow \{6\mbox{-}j\}_{1}\; \mbox{\scriptsize{\&}}\; \{6\mbox{-}j\}_{2} ,\\
\{6\mbox{-}j\}_{1}\stackrel{{\cal{R}}_{egge}^{*}}\longrightarrow \{6\mbox{-}j\}_{0}\; \mbox{\scriptsize{\&}}\; \{6\mbox{-}j\}_{2},\\
\{6\mbox{-}j\}_{2}\stackrel{{\cal{R}}_{egge}^{*}}\longrightarrow \{6\mbox{-}j\}_{0}\; \mbox{\scriptsize{\&}}\; \{6\mbox{-}j\}_{1}.
\end{array}
\end{equation}
The order in right equation members above is not relevant, one can take $1,2$ or $2,1$ and so on.\\
Clearly the set ${\tt{S}}_4 (2)$ is {\tt stable} under ${\cal{R}}_{egge}^{*}$. 
The number of its elements is {\small Card}(${\tt{S}}_4 (2))$ i.e. $\#({\tt{S}}_4 (2))= 3\times 24=72$. Now the main task is to discover what are the parameters which define
the presumed partitions ${\tt{S}}_4 (0)\cdots {\tt{S}}_4 (5)$.\\ [0.2em]
{\tt After solving the partitions, the results are summarized below: }\\ [0.1em]
Here we implicitly assume that $l \in [1,3]$ and $l,m,n$ cyclic on $[1,2,3]$, even if there is another index like  $l'$.
A notation like (all $p\neq$) means that for the full set of $p_1,p_2,p_3,p_4$ all  $p$ are different between them.
\begin{equation}\label{eq:S40}
\begin{array}{cc}
{\tt{S}}_4 (0)=\big \{\{6\mbox{-}j\}\big\}\,|&\,  \mbox{(all $q$ equal )}\\
&\mbox{{\tt or}}\\
&(q_3\neq  q_1)\,{ \tt{and}}\, ((\mbox{all $p$ equal})\,{ \tt{or}}\,(\mbox{all $p_l$ equal $\neq p_4$}))\\
&\mbox{{\tt or}}\\
 &(q_3\neq  q_1)\,{ \tt{and}}\,(p_1=p_2=p_4\neq p_3 )
\end{array}.
\end{equation}
\begin{equation}\label{eq:S41}
\begin{array}{cc}
{\tt{S}}_4 (1)=  \big\{\{6\mbox{-}j\}\big\}\,|& (q_l =  q_m\neq q_n ) \,{\tt and}\,(\mbox{all }p_{l'}\neq p_4)\,{ \tt{and}}
\, (p_{l'}=p_{m'}\neq p_{n'})\\
&\mbox{{\tt or}}\\
&(q_l =  q_m \neq q_n) \,{\tt and}\,(\mbox{all }\,p_{l'} \neq)
 \mbox{ \tt{and}}\;((p_1=p_4) \mbox{ \tt{or}}\,(p_2=p_4))\\
&\mbox{{\tt or}}\\
&(q_l =  q_m \neq q_n) \,{\tt and}\, (p_1 \neq p_2) \,{\tt and}\, (p_2=p_3)\,{\tt and}\, (p_1=p_4)\\
&\mbox{{\tt or}}\\
&( \,q_2 =  q_3\neq q_1 ) \,{\tt and}\, (p_1 \neq p_2) \,{\tt and}\,(p_1=p_3)\,{\tt and}\, (p_2=p_4)\\
&\mbox{{\tt or}}\\
&( \,q_1 =  q_2\neq q_3 ) \,{\tt and}\,( p_1=p_2\neq p_3)\,{\tt and} \,(p_3=p_4 )
\end{array}.
\end{equation}
\begin{eqnarray}\label{eq:S42}
\begin{array}{cc}
{\tt{S}}_4 (2)= \big\{\{6\mbox{-}j\}\big\}\,|&\;( q_l =  q_m )\,{\tt{and}}\;(\mbox{all}\; p  \neq )\\
&\mbox{{\tt or}}\\
&(\mbox{all}\; q  \neq )\,{\tt{and}}\;(p_l =  p_m )\,{\tt{and}}\;(\mbox{all}\; p_l  \neq p_4 ) \\
&\mbox{{\tt or}}\\
&(\mbox{all }q  \neq )\,{\tt{and}}\;( p_l =  p_m )\,{\tt{and}}\;(p_n=p_4 ) \\
&\mbox{{\tt or}}\\
&(\mbox{all}\; q  \neq )\,{\tt{and}}\;(\mbox{all}\;p_l \neq  )\,{\tt{and}}\;((p_1=p_4 )\,{\tt{or}}\;(p_2=p_4 )) 
\end{array}.
\end{eqnarray}
\begin{equation}\label{eq:S403}
\begin{array}{cc}
{\tt{S}}_4 (3) = & \emptyset 
\end{array}.
\end{equation}
\begin{equation}\label{eq:S404}
\begin{array}{cc}
{\tt{S}}_4 (4)=& \emptyset 
\end{array}.
\end{equation}
\begin{eqnarray}\label{eq:S45}
\begin{array}{cc}
{\tt{S}}_4 (5)= &\big\{\{6\mbox{-}j\}\big\}\;|\; (\mbox{all }q \neq ) \,  {\tt{and}}\,  
(\mbox{all }p \neq ).
\end{array}
\end{eqnarray}

${\tt{S}}_4 (0)$ is {\tt stable} under  ${\cal{R}}_{egge}^{*}$. No new triangle is created,
consequence of (\ref{eq:Alteq1}-\ref{eq:Alteq4}).
Actually each of  ${\tt{S}}_4 (1), {\tt{S}}_4 (2), {\tt{S}}_4 (5)$ are {\tt stable} also. Our notion of 'stability' for these sets has been well defined.
It is known also that [example of the matrices $\mathbf{R}_\kappa$] any eigenspace is stable under $\mathbf{R}_\kappa$.
Is there a correlation between these eigenspaces and sets  ${\tt{S}}_4 (\,)$? It would be worth to deepen in this topic. \\
\hspace*{1em}Moreover why our initial partitions imply that ${\tt{S}}_4 (3)$ and ${\tt{S}}_4 (4)$ are empty sets arises from our natural definitions (\ref{eq:DisjtEqs})
and the fact that additional constraint of inequalities to (\ref{eq:S42}) would lead to 
${\tt{S}}_4 (3)\rightarrow {\tt{S}}_4 (4)\Rightarrow {\tt{S}}_4 (3)\equiv {\tt{S}}_4 (4) \rightarrow {\tt{S}}_4 (5)\Rightarrow
{\tt{S}}_4 (3)\equiv{\tt{S}}_4 (4)\equiv {\tt{S}}_4 (5)$. Then ${\tt{S}}_4 (3),  {\tt{S}}_4 (4)$ cannot exist.
That can be understood only after solving the algebraic partitions.\\
Initial identification  of algebraic features of ${\tt{S}}_4 (0)$ was easily found by hand calculation, however, for the remaining sets, it becomes less obvious.
That is why it was decided to write a program ({\tt symmetryregge}) called for help, comparing a lot of ${\tt{S}}_4 $ sets. The process is 
not fully automatic and requires careful analysis. Originally scheduled for  $6$-$j^{S}$  super-symbols, it turns out that it obviously applies to standard $6$-$j$ symbols,
where no distinction has to be made between parities $\alpha, \beta, \gamma$. In this case we only have to handle $\alpha$. \\
Our logical partitions of the disjoint subsets  themselves, i.e. by using {\tt{'and'}} and {\tt{'or'}},
are quite right, but maybe not necessarily the best, because a general symmetry does not appear since sometimes $q_1,q_3,p_4$ are singularized. 
It should be possible to find other rearrangements.  \\ 
\hspace*{1em}Now all features of standard $6$-$j$ symbols can be symbolically summarized by a sequence, where over each subset is indicated its cardinal \#: 
\begin{equation}\label{eq:ReggeforStandard6j}
\{6\mbox{-}j\}\stackrel{{\cal{R}}_{egge}^{*}}\longrightarrow \;\,
 \stackrel{\! \#=24}{\tt{S}_4 (0)}
\oplus \; \stackrel{\! \#=48}{\tt{S}_4 (1)}
\oplus \; \stackrel{\! \#=72}{\tt{S}_4(2)}
\oplus \; \stackrel{\! \#=144}{\tt{S}_4 (5)}.
\end{equation}
Thus adding Regge symmetries to standard $6$-$j$ symbols  changes their symmetry groups according to the subset to which they belong.
The table below shows the related isomorphisms.
\begin{equation}\label{eq:Isomorphism}
\begin{array}{|c|c|c|c|}
\multicolumn{4}{c}{\{6$-$j\}+ \mbox{Regge symmetry}}\\ [0.1em] \hline
\;{\tt{S}_4 (0)}\; & {\tt{S}_4 (1)}& \;{\tt{S}_4 (2)}\; & \;{\tt{S}_4 (5)}\; \\ \hline
S_4  & \; S_4 \times S_2 \; & \; S_4 \times (S_3/S_2)\;  & \; S_4 \times S_3 \; \\ \hline
\end{array}
\end{equation}

				\section{Regge symmetry of $6$-$j^{S}$ symbols }\label{ReggeSect4}
\renewcommand{\theequation}{\ref{ReggeSect4}.\arabic{equation}}
\setcounter{equation}{0}
\hspace*{1em}In 1992 was published a preliminary study dealing with this subject \cite{Daumensetal0} and  'appended' inside the paper which followed \cite {Daumensetal1}. 
The authors then were able to announce unexpected results namely:
a similar Regge symmetry exists also for  $6$-$j^{S}$ symbols \cite{Daumensetal1}, but sometimes smaller than in the case of $so(3)$ \cite{Daumensetal0}.
Their proof, self-qualified of {\it ''laborious''} was omitted.\\
Results  were presented by using phase factors, with inference to Bargmann's representation \cite {Bargmann}.\\
\hspace*{1em}As since 1993, nothing was known about partition properties, it seemed interesting to further develop this topic  also for $6$-$j^{S}$ symbols,
as a complement to  our previous work \cite {L.B.Nuov1.2006}.\\
In the present case, it will be seen that use of parities $\pi=\alpha,\beta,\gamma$  and $q$ formulation provides clear and immediate solutions,
{\tt free of explicit phase factors and without need of another representation.}
General results cited just above \cite{Daumensetal0, Daumensetal1} are significantly better understood  by means of our approach with partitions. 
Moreover it brings unknown information related to $ p $ and $ q $.\\
\hspace*{1em}First of all, in virtue of our analytic formulas for $6$-$j^{S}$ symbols,  (\ref{eq:General6jS1})
compared to (\ref{eq:General6j1}), clearly one can assert that 
a $6$-$j^{S}$, as well as a standard $6$-$j$,  possesses the same symmetry $S_4$. This property can thus be used as desired.\\
In this section, the only difference lies in  the fact that triangles $p$ may be integer or half-integer as well as quadrangles $q$. 
Depending on parities $\pi=\alpha, \beta, \gamma$, one has the following properties:\\
$\pi=\alpha$ or $\gamma$ $\Rightarrow$ all $q$ are integers.\\
$\pi=\beta$ $\Rightarrow$ $\exists \,! \,q_{l_{}^{\stackrel{*}{}} }$ integer (the other two are half-integer). $ l_{}^{\stackrel{*}{}}$ is a distinguished index.\\
As any $q$ half-integer is excluded \footnote{ Spins like$\begin{array}{c}(2n+1)/4\end{array}$ forbidden.}
 from the Regge transformations the conclusion is clear:\\
Case $\boldsymbol{\alpha}$ or $\boldsymbol{\gamma}$: all Regge symmetries are valid. \\
Case $\boldsymbol{\beta}$: a single transformation is valid. \\
After noting that parity $\pi$ remains invariant under any  Regge transformation, one can write:
\begin{equation}
\scriptsize
\left\{ \begin{array}{ccc} J_1 &J_2 & J_3 \\ j_1 & j_2 & j_3 \end{array} \right\}_{\!\boldsymbol{\alpha},\boldsymbol{\gamma}}^{S}=
\left\{ \begin{array}{ccc} J_l &\frac{1}{2}q_l-J_m & \frac{1}{2}q_l-J_n  \\ [0.1em]
j_l &\frac{1}{2}q_l-j_m &\frac{1}{2}q_l-j_n \end{array} \right\}_{\!\boldsymbol{\alpha},\boldsymbol{\gamma}}^{S} \normalsize
 \quad\;  l \in[1,3] \;   l,m,n \;  \mbox{\hspace{0.1em}cyclic\;on  } \hspace{0.1em}[1,2,3] ,
\end{equation} 
\begin{eqnarray}
\scriptsize
\left\{ \begin{array}{ccc} J_1 &J_2 & J_3 \\ j_1 & j_2 & j_3 \end{array} \right\}_{\!\boldsymbol{\alpha},\boldsymbol{\gamma}}^{S}&=
\scriptsize
\left\{ \begin{array}{ccc}\frac{1}{2}q_l-j_m  &\frac{1}{2}q_m-j_n & \frac{1}{2}q_n-j_l  \\ [0.1em]
\frac{1}{2}q_l-J_m  &\frac{1}{2}q_m-J_n & \frac{1}{2}q_n-J_l   \end{array} \right\}_{\!\boldsymbol{\alpha},\boldsymbol{\gamma}}^{S}&=
\scriptsize
 \left\{ \begin{array}{ccc}\frac{1}{2}q_l-j_n  &\frac{1}{2}q_m-j_l & \frac{1}{2}q_n-j_m  \\ [0.1em]
\frac{1}{2}q_l-J_n  &\frac{1}{2}q_m-J_l & \frac{1}{2}q_n-J_m   \end{array} \right\}_{\!\boldsymbol{\alpha},\boldsymbol{\gamma}}^{S}.
\end{eqnarray}

\begin{equation}
\scriptsize
\left\{ \begin{array}{ccc} J_{l_{}^{\stackrel{*}{}} }&J_{m } & J_{n}\\ j_{l_{}^{\stackrel{*}{}}}& j_{m}  & j_{n} \end{array} \right\}_{\boldsymbol{\beta}}^{S}\,=
\left\{ \begin{array}{ccc} J_{l_{}^{\stackrel{*}{}} } &\frac{1}{2}q_{l_{}^{\stackrel{*}{}}}-J_{m}  & \frac{1}{2}q_{l_{}^{\stackrel{*}{}}}-J_{n}  \\ [0.1em]
j_{l_{}^{\stackrel{*}{}}} &\frac{1}{2}q_{l_{}^{\stackrel{*}{}}}-j_{m}  &\frac{1}{2}q_{l_{}^{\stackrel{*}{}}}-j_{n} \end{array} \right\}_{\boldsymbol{\beta}}^{S}\normalsize
\quad \mbox{(only one relation for $\boldsymbol{\beta}$)}.
\end{equation}

				\subsection{Features of $6$-$j^{S}$ symbols generated by Regge transformations}\label{ReggeSect4prime}
\renewcommand{\theequation}{\ref{ReggeSect4prime}\alph{equation}}
\setcounter{equation}{0}
It is obvious that, for parity $\boldsymbol{\alpha}$ or $\boldsymbol{\gamma}$, properties like (\ref{eq:S40}-\ref{eq:S45}) are still valid. Thus
\begin{equation}
\{6\mbox{-}j\}_{\alpha,\gamma}^{S}\stackrel{{\cal{R}}_{egge}^{*}}\longrightarrow \;
 \stackrel{\! \#=24}{\tt{S}_4 ^{\it S}(0)}
\oplus \; \stackrel{\! \#=48}{\tt{S}_4 ^{\it S}(1)}
\oplus \; \stackrel{\! \#=72}{\tt{S}_4 ^{\it S}(2)}
\oplus \; \stackrel{\! \#=144}{\tt{S}_4 ^{\it S}(5)}.
\end{equation}

However for parity $\pi=\boldsymbol{\beta}$, only sets like ${\tt{S}}_4  ^{\it S}(0)$  or ${\tt{S}}_4  ^{\it S}(1)$ can exist. They are defined by
\begin{equation}\label{eq:S40beta}
{\tt{S}_4 ^{\it S}(0)}= \,\left\{ \{6\mbox{-}j\}^{S}\right\}\;| \,(\overline{q}= \overline{q}') \,{\tt or}
\,(\mbox{($\overline{q}\neq \overline{q}'$) \tt{and}}\, ((p=p')\,{\tt  or}\,(\overline{p}= \overline{p}'))).
\end{equation}
\begin{equation}\label{eq:S41beta}
{\tt{S}_4 ^{\it S}(1)}= \,\left\{ \{6\mbox{-}j\}^{S}\right\}\;| \,(\overline{q}\neq \overline{q}') \,{\tt and}\;(p\neq p')
\,{\tt and}\;(\overline{p}\neq \overline{p}').
\end{equation}
One can then write 
\begin{equation}
\{6\mbox{-}j\}_{\beta}^{S}\stackrel{{\cal{R}}_{egge}^{*}}\longrightarrow \;
 \stackrel{\! \#=24}{\tt{S}_4 ^{\it S}(0)}
\oplus \; \stackrel{\! \#=48}{\tt{S}_4 ^{\it S}(1)}.
\end{equation}
Analogous isomorphism tables to (\ref{eq:Isomorphism}) are as follows:
\begin{equation}\label{eq:IsomorphismSag}
\begin{array}{|c|c|c|c|}
\multicolumn{4}{c}{\{6\mbox{-}j\}_{\alpha,\gamma}^{S}+ \mbox{Regge symmetry}}\\ [0.2em] \hline
\;{\tt{S}_4 ^{\it S}(0)}\; & \;{\tt{S}_4 ^{\it S}(1)}\; & \;{\tt{S}_4 ^{\it S}(2)}\; & \;{\tt{S}_4 ^{\it S}(5)}\; \\ \hline
S_4  & \; S_4 \times S_2 \; & \; S_4 \times (S_3/S_2)\;  & \; S_4 \times S_3 \; \\ \hline
\end{array}
\end{equation}
and
\begin{equation}\label{eq:IsomorphismSb}
\begin{array}{|c|c|}
\multicolumn{2}{c}{\{6\mbox{-}j\}_{\beta}^{S}+ \mbox{Regge symmetry}}\\ [0.2em] \hline
\;{\tt{S}_4 ^{\it S}(0)}\; & \;{\tt{S}_4 ^{\it S}(1)}\\ \hline
\; S_4  \; & \; S_4 \times S_2 \\ \hline
\end{array}
\end{equation}
Our conclusion does not differ appreciably from that of Daumens {\it et al.} \cite{Daumensetal0,Daumensetal1}, 
but is much more accurate with unknown information that we have highlighted.
							\section{Program notes}\label{TableRotenberg}
Mathematical properties  detailed in  \bf \ref{SuperFormula}  \rm has made easier the coding of a computing program ({\tt superspins}), written in fortran 95.
As  a result, editing the table of $6$-$j^{S}$ symbols, from  
\scriptsize $\left\{ \begin{array}{ccc} 0 & 0 & 0\\ 0 & 0 & 0 \end{array} \right\}^{S}$\normalsize to 
\scriptsize $\left\{ \begin{array}{ccc} 10 & 10 & 10 \\ 10 & 10 & 10 \end{array} \right\}^{S}$\normalsize 
takes only 6 seconds with  a small home computer. Of course this was processed in the same spirit than that used for producing
the famous Rotenberg's tables \cite{Rot}. All is computed with integers, factorial and prime numbers. Redundancies were avoided as far as possible.
As expected, the fit with the reduced data published in Ref. \cite{Daumensetal1} is perfect.\\ We have enlarged the basis of prime numbers:
$2$ to $31$ becomes $2$ to $53$, namely  $\tt p_{1}$ to  $\tt p_{16}$. Program ({\tt superspins}) computes their exponents $\tt e_i,  i \in[1,16]$.
You have to evaluate the expression- generally fractional- $\prod_{i=1}^{16}{\tt (p_{i})}^{ \tt e_i}$, take its square root and multiply by the integer which follows the ampersand
\scriptsize  \bf $\&$\rm \normalsize.
If beyond $53$, this integer is showed under the form of a prime numbers product. Then you obtain the exact numerical value of a $6$-$j^{S}$ symbol.\\
 In addition, the parity is listed as \scriptsize \bf  $<$a$>$, $<$b$>$, $<$g$>$ \rm \normalsize respectively for $\alpha, \beta, \gamma$ .
Excerpts of the table are given below.\\[0.1em]

\includegraphics [width=\textwidth]{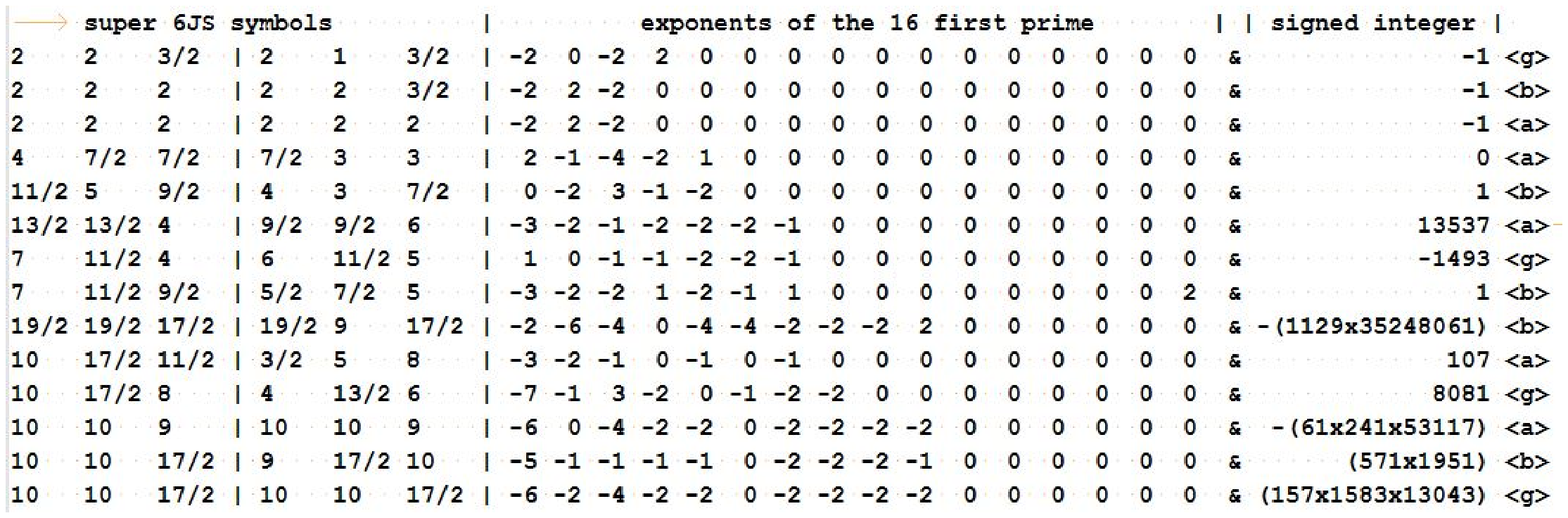}\\ 
Output files (format .txt): \\
{\tt superspinstest,superspzeroa,superspzerob,superspzerog,}\\
{\tt symmreggetesta0,symmreggetesta1,symmreggetesta2,symmreggetesta5,}\\
{\tt symmreggetestb0,symmreggetestb1,}\\
{\tt symmreggetestg0,symmreggetestg1,symmreggetestg2,symmreggetestg5.}\\
The first contains the full table, the other file names suggest their contents.\\
Of course a slight reduction and change of ({\tt superspins})$\rightarrow$ ({\tt standardsymbols}) allows us to recover instantly 
the part of  Rotenberg's table \cite{Rot} which lists the standard $6$-$j$ symbols.

\section{Conclusion}
\hspace*{1em}Contrary to received ideas, first of all it seems improper to assert that adding Regge symmetry to any $6$-$j$ symbols  
automatically gives them a symmetry group isomorphic to $S_4\times S_3$, because subsets like ${\tt{S}_4 (0)},{\tt{S}_4 (1)},{\tt{S}_4 (2)}$ 
are quite well defined, according to Table (\ref{eq:Isomorphism}).
The same holds for any super $6$-$j^S$ symbols with the subsets ${\tt{S}_4 ^{\it S}(0)},{\tt{S}_4 ^{\it S}(1)},{\tt{S}_4 ^{\it S}(2)}$,
 according to Tables (\ref {eq:IsomorphismSag}),(\ref{eq:IsomorphismSb}).
In the same way the assertion that {\it``the symmetry group of  an arbitrary $6$-$j^{S}$ symbol contains at least $48$ elements''}
\cite{Daumensetal1} should be re-formulated. \\
In summary the possible isomorphic symmetry groups are the following:
\begin{equation}
\begin{array}{|c|c|c|c|}
\multicolumn{4}{c}{\{6\mbox{-}j\}\;\mbox{and }\,\{6\mbox{-}j\}_{\alpha,\gamma}^{S}}\\ [0.2em] \hline
S_4  & \; S_4 \times S_2 \; & \; S_4 \times (S_3/S_2)\;  & \; S_4 \times S_3 \; \\ \hline
\end{array}
\qquad
\begin{array}{|c|c|}
\multicolumn{2}{c}{\{6\mbox{-}j\}_{\beta}^{S}}\\ [0.2em] \hline
\; S_4  \; & \; S_4 \times S_2 \\ \hline
\end{array}
\end{equation}
\hspace*{1em}This study made us discover a hidden classification  of each $6$-$j$ symbol, according to their belonging to four disjoint sets.
A new notation could be $_{c}\{6\mbox{-}j\}$ where \scriptsize$c=0,1,2,5$\normalsize. These symbols are directly evidenced in spectroscopic experiments
 {\tt in nuclear, atomic or molecular physics}, generally for evaluating transition intensities or interaction energy. 
As a first step, a review of the type of  implied symbols could reveal whether parameter {\scriptsize c} is significant or not.


\end{flushleft}
\end{document}